\RequirePackage{ifpdf}
\documentclass[hyper,letterpaper]{JHEP3}
\usepackage{amsmath,amssymb,amsfonts,mathbbol}
\usepackage{fancybox}
\usepackage{graphicx, wrapfig}
\usepackage{multirow}
\usepackage{verbatim}
\usepackage{appendix}
\usepackage{slashed}
\usepackage{url}
\usepackage{float}
\usepackage{cite}
\DeclareSymbolFontAlphabet{\amsmathbb}{AMSb}

\title{A resurgence analysis of the $SU(2)$ Chern-Simons partition functions on a Brieskorn homology sphere $\Sigma(2,5,7)$}

\author{Sungbong Chun \\
Walter Burke Institute for Theoretical Physics, California Institute of Technology, Pasadena, CA 91125 USA}

\abstract{We perform a resurgence analysis of the $SU(2)$ Chern-Simons partition function on a Brieksorn homology sphere $\Sigma(2,5,7)$. Starting from an exact Chern-Simons partition function, we study the Borel resummation of its perturbative expansion. 
\\
\\
\\
\\
\\
\\
\\
{\tt CALT-TH-2017-003}}

\begin{document}
\cornersize{1}

\section{Introduction}
We perform a resurgence analysis of the $SU(2)$ Chern-Simons partition function on a Bireksorn homology sphere, following \cite{GMP}. Consider the Chern-Simons action with a gauge group $G$ on a 3-manifold $M_{3}$:
$$CS(A) = \frac{1}{8\pi^{2}} \int_{M_{3}} A \wedge dA + \frac{2}{3}A \wedge A \wedge A,$$
where $A$ is a Lie algebra (ad$G$) valued 1-form on $M_{3}$. Classical solutions of this action are the flat connections, satisfying $F_{A} = dA + A \wedge A = 0$. The Chern-Simons partition function at level $k$ can be expanded with a perturbation parameter $1/k$, around the flat connections:

\begin{equation}
Z_{CS}(M_{3}) = \sum_{\alpha \in \mathcal{M}_{\text{flat}}(M_{3}, G)} e^{2 \pi i k CS(\alpha)}Z^{\text{pert}}_{\alpha}.
\label{eqn:pert}
\end{equation}
Above, $\mathcal{M}_{\text{flat}}(M_{3}, G)$ is the moduli space of flat $G$-connections on $M_{3}$, and we have assumed a discrete moduli space. When $k$ is an integer, $CS(A)$ is only defined modulo 1. 

The exact partition function $Z_{CS}(M_{3}) = \int \mathcal{D}A e^{2 \pi i k CS(A)}$ can be recovered from its perturbative expansion by a resurgence analysis of Jean \'{E}calle \cite{Ecalle}. We first analytically continue $k$ to compex values and apply the method of steepest descent. Then, perform a Borel transformation and resummation of the perturbative partition function, to recover the exact partition function. Surprisingly, the exact partition function is now written as a linear sum of the ``homological blocks'' \cite{GMP}:
\begin{equation}
Z_{CS}(M_{3}) = \sum_{a \, \text{abelian}} e^{2 \pi i k CS(\alpha)}Z_{a}.
\label{eqn:abelianDecomposition}
\end{equation}
Above, $Z_{a}$ gets contributions from \textit{both} the abelian flat connection $a$ and the irreducible flat connections. In \cite{GPV}, it was proposed that the partition function in this form allows a ``categorification,'' in a sense that it is a ``S-transform'' of a vector whose entries are integer-coefficient Laurent series in $q = e^{2\pi i /k}$.

In this paper, we provide a supporting example of \cite{GMP}. First, we perform a resurgence analysis of $SU(2)$ Chern-Simons partition function on  a Brieskorn homology sphere, $M_{3} = \Sigma(2,5,7)$. We start with the exact partition function $Z_{CS}(\Sigma(2,5,7))$, which is written as a linear sum of ``mock modular forms'' \cite{HikamiBrieskorn}. Then, we consider its perturbative expansion and perform a Borel resummation. The Borel resummation in effect recovers the full partition function $Z_{CS}(\Sigma(2,5,7))$, and we observe a Stokes phenomenon which encodes the non-perturbative contributons to the partition function. 

\section{Setups for the Borel resummation in Chern-Simons theory}
In this section, we provide necessary notations and setups for the Borel resummation in Chern-Simons theory. A complete and concise review can be found in section 2 of \cite{GMP}. 

Let us start with the exact Chern-Simons partition function $Z_{CS}(M_{3}) = \int \mathcal{D}A e^{2 \pi i k CS(A)}$, integrated over $G=SU(2)$ connections. Next, analytically continue $k$ to complex values and apply the method of steepest descent on the Feynman path integral \cite{Marino, Garoufalidis, GLM, ArgyresUnsal, Kashani-Poor, CostinGaroufalidis, WittenAnalytic, KontsevichPerimeter,KontsevichSCGP,KontsevichTFC}. Then, the integration domain is altered to a middle-dimensional cycle $\Gamma$ in the moduli space of $G_{\mathbb{C}} = SL(2,\mathbb{C})$ connections, which is the union of the steepest descent flows from the saddle points. To elaborate, the moduli space is the universal cover of the space of $SL(2,\mathbb{C})$ connections modulo ``based'' gauge transformations, in which the gauge transformations are held to be $1$ at the designated points. In sum, the partition function becomes:
\begin{equation}
Z_{CS}(M_{3}) = \int_{\Gamma} \mathcal{D}A e^{2 \pi i k CS(A)}, \quad k \in \mathbb{C}.
\label{eqn:ZoverGamma}
\end{equation}

\subsection{Borel resummation basics}
Partition function of form Equation \ref{eqn:ZoverGamma} is interesting, for its perturbative expansion can be regarded as a \textit{trans-series} expansion, which can be Borel resummed. Let us provide here the basics of Borel resummation, following \cite{Marino}. The simplest example of a trans-series is a formal power series solution of Euler's equation:
$$\frac{d \varphi}{dz} + A \varphi(z) = \frac{A}{z}, \quad \varphi_{0}(z) = \sum_{n \geq 0} \frac{A^{-n}n!}{z^{n+1}}.$$
One may view the above trans-series as a perturbative (in $1/z$) solution to the differential equation, but the solution has zero radius of convergence. By the Borel resummation, however, one can recover a convergent solution. When a trans-series is of form $\varphi(z) = \sum_{n \geq 0} a_{n}/z^{n}$ with $a_{n} \sim n!$, its Borel transformation is defined as:
$$\hat{\varphi}(\zeta) = \sum_{n \geq 1} a_{n} \frac{\zeta^{n-1}}{(n-1)!}.$$
The Borel transformation $\hat{\varphi}(\zeta)$ is analytic near the origin of $\zeta$-plane. If we can analytically continue $\hat{\varphi}(\zeta)$ to a neighborhood of the positive real axis, we can perform the Laplace transform:
$$S_{0}\varphi(z) = a_{0} + \int_{0}^{\infty} e^{- z \zeta}\hat{\varphi}(\zeta) d \zeta,$$
where the subscript ``0'' indicates that the integration contour is along the positive real axis, $\{ \arg(z) = 0 \}$. It can be easily checked that the asymptotics of the above integral coincides with that of $\varphi(z)$. When $S_{0}\varphi(z)$ converges in some region in the $z$-plane, $\varphi(z)$ is said to be Borel summable, and $S_{0}\varphi(z)$ is called the Borel sum of $\varphi(z)$. 

\subsection{Chern-Simons partition function as a trans-series}

Saddle points of the Chern-Simons action form the moduli space of flat connections $\tilde{M}$,
whose connected components $\tilde{M}_{\tilde{\alpha}}$ are indexed by their ``instanton numbers,''
$$\tilde{\alpha} = (\alpha, CS(\tilde{\alpha})) \in \mathcal{M}_{\text{flat}}(M_{3},SL(2, \mathbb{C})) \times \mathbb{Z}.$$ 
Here, $CS(\tilde{\alpha})$ denotes the value of Chern-Simons action at $\alpha$, without moding out by 1. Following \cite{GMP}, we will call a flat connection \textit{abelian} (\textit{irreducible}, resp.), if the stabilizer is $SU(2)$ or $U(1)$ ($\{ \pm 1\}$, resp.) action on  $Hom(\pi_{1}(M_{3}),SU(2))$. 

Now, let $\Gamma_{\tilde{\alpha}}$ be the union of steepest descent flows in $\tilde{M}$, starting from $\tilde{\alpha}$. The integration cycle $\Gamma$ is then given by a linear sum of these ``Lefshetz thimbles.''

\begin{equation}
\Gamma = \sum_{\tilde{\alpha}} n_{\tilde{\alpha},\theta}\Gamma_{\tilde{\alpha},\theta},
\label{eqn:GammaDecomposition}
\end{equation}
where $\theta = \arg(k)$, and $n_{\tilde{\alpha},\theta} \in \mathbb{Z}$ are the \textit{trans-series} parameters, given by the pairing between the submanifolds of steepest descent and ascent. The value of $\theta$ is adjusted so that there is no steepest descent flow between the saddle points. Let $I_{\tilde{\alpha},\theta}$ be the contribution from a Lefshetz thimble $\Gamma_{\tilde{\alpha},\theta}$ to $Z_{CS}(M_{3})$ in Equation \ref{eqn:ZoverGamma}:

$$I_{\tilde{\alpha},\theta} = \int_{\Gamma_{\tilde{\alpha},\theta}} \mathcal{D}A e^{2 \pi i k CS(A)},$$
which can be expanded in $1/k$ near $\tilde{\alpha}$ as:
$$I_{\tilde{\alpha},\theta} \sim e^{2 \pi i k CS(\tilde{\alpha})}Z^{\text{pert}}_{\alpha}, \quad \text{where} \quad Z^{\text{pert}}_{\alpha} = \sum_{n=0}^{\infty} a_{n}^{\alpha}k^{-n+(d_{\alpha}-3)/2}, \quad d_{\alpha} = dim_{\mathbb{C}}\tilde{\mathcal{M}}_{\tilde{\alpha}}.$$
In sum, we can write the Chern-Simons partition function in the form:

\begin{equation}
Z_{CS}(M_{3};k) = \sum_{\tilde{\alpha}} n_{\tilde{\alpha},\theta}I_{\tilde{\alpha},\theta} \sim \sum_{\tilde{\alpha}} n_{\tilde{\alpha},\theta}e^{2 \pi i k CS(\tilde{\alpha})}Z^{\text{pert}}_{\alpha}(k),
\label{eqn:trans-series}
\end{equation}
which is a trans-series expansion of the Chern-Simons partition function. From the asymptotics given by this trans-series, we can apply Borel resummation and recover the full Chern-Simons partition function. Note that Equation \ref{eqn:trans-series} depends on the choice of $\theta = \arg(k)$. In fact, as we vary $\theta$, the value of $I_{\tilde{\alpha},\theta}$ jumps to keep the whole expression continuous in $\theta$ as follows:
\begin{equation}
I_{\tilde{\alpha},\theta_{\tilde{\alpha}\tilde{\beta}}+\epsilon} = I_{\tilde{\alpha},\theta_{\tilde{\alpha}\tilde{\beta}}-\epsilon} + m_{\tilde{\alpha}}^{\tilde{\beta}}I_{\tilde{\beta},\theta_{\tilde{\alpha}\tilde{\beta}}-\epsilon}.
\label{eqn:Stokes}
\end{equation}
This is called the Stokes phenomenon, and it happens near the Stokes rays $\theta = \theta_{\tilde{\alpha}\tilde{\beta}} \equiv \frac{1}{i}\arg(S_{\tilde{\alpha}}-S_{\tilde{\beta}})$. The trans-series parameters $n_{\tilde{\alpha},\theta}$ jump accordingly to keep $Z_{CS}(M_{3};k)$ continuous in $\theta$. The coefficients $m_{\tilde{\alpha}}^{\tilde{\beta}}$ are called Stokes monodromy coefficients.

\section{Exact partition function $Z_{CS}(\Sigma(2,5,7))$}
\label{sec:exact}
Before going into the resurgence analysis of $Z_{CS}(\Sigma(2,5,7))$, let us provide here the exact partition function $Z_{CS}(\Sigma(2,5,7))$. We first compute the Witten-Reshetikhin-Turaev (WRT) invariant $\tau_{k}(\Sigma(p_{1},p_{2},p_{3}))$ and then write the exact $SU(2)$ Chern-Simons partition function in terms of WRT invariants as follows:
\begin{equation}
Z_{CS}(\Sigma(p_{1},p_{2},p_{3})) = \frac{\tau_{k}(\Sigma(p_{1},p_{2},p_{3}))}{\tau_{k}(S^{2} \times S^{1})}.
\label{eqn:WRTandCSpartition}
\end{equation}
Here, $k$ is the level of Chern-Simons theory.\footnote{To be more precise, $k$ must be replaced by $k+2$. However, our interest in this paper is to recover the full partition function from a perturbative expansion in $1/k$. Therefore, we will assume $k$ to be large, and replace $k+2$ with $k$ here.}

WRT invariants for Seifert homology spheres can be computed from their surgery presentations \cite{LawrenceRozansky}. In this paper, we focus on a specific type of Seifert homology spheres, the so-called Bireskorn homology spheres. A Brieskorn manifold $\Sigma(p_{1},p_{2},p_{3})$ is defined as an intersection of a complex unit sphere $|z_{1}|^{2}+|z_{2}|^{2}+|z_{3}|^{2}=1$ and a hypersurface $z_{1}^{p_{1}}+z_{2}^{p_{2}}+z_{3}^{p_{3}}=0$. When $p_{1},p_{2},p_{3}$ are coprime integers, $\Sigma(p_{1},p_{2},p_{3})$ is a homology sphere with three singular fibers. From the surgery presentation of $\Sigma(p_{1},p_{2},p_{3})$, we can write its WRT invariant, which can be written a linear sum of mock modular forms \cite{HikamiBrieskorn, LawrenceZagier}. In particular, when $1/p_{1}+1/p_{2}+1/p_{3} < 1$, we can write:
\begin{equation}
e^{\frac{2 \pi i}{k}(\frac{\phi(p_{1},p_{2},p_{3})}{4}-\frac{1}{2})}(e^{\frac{2 \pi i}{k}} -1) \tau_{k}(\Sigma(p_{1},p_{2},p_{3})) = \frac{1}{2}\tilde{\Psi}_{p_{1}p_{2}p_{3}}^{(1,1,1)}(1/k).
\label{eqn:WRTmodular}
\end{equation}
Let us decode Equation \ref{eqn:WRTmodular}. First of all, $\tau_{k}(\Sigma(p_{1},p_{2},p_{3}))$ is the desired WRT invariant, normalized such that $\tau_{k}(S^{3}) = 1$ and $\tau_{k}(S^{2} \times S^{1}) = \sqrt{\frac{k}{2}}\frac{1}{\sin(\pi / k)}.$ Next, the number $\phi(p_{1},p_{2},p_{3})$ is defined as:
\begin{gather}
\phi(p_{1},p_{2},p_{3}) = 3 - \frac{1}{p_{1}p_{2}p_{3}} + 12 (s(p_{1}p_{2},p_{3})+s(p_{2}p_{3},p_{1})+s(p_{3}p_{1},p_{2})), \nonumber \\
\text{where} \quad s(a,b) = \frac{1}{4b}\sum_{n=1}^{b-1}\cot(\frac{n \pi}{b})\cot(\frac{n a \pi}{b}). \nonumber
\end{gather}
Finally, $\tilde{\Psi}_{p_{1}p_{2}p_{3}}^{(1,1,1)}$ is a linear sum of mock modular forms $\tilde{\Psi}_{p_{1}p_{2}p_{3}}^{a}$, namely:
\begin{gather}
\tilde{\Psi}_{P}^{a}(1/k) = \sum_{n \geq 0} \psi_{2P}^{a}(n)q^{n^{2}/4P}, \quad \text{where} \quad \psi_{2P}^{a}(n) = \begin{cases}
\pm 1 & \text{$n \equiv \pm a$ mod $2P$} \\   0 & \text{otherwise}
\end{cases} \label{eqn:modular} \\
\tilde{\Psi}_{p_{1}p_{2}p_{3}}^{(1,1,1)}(1/k) = -\frac{1}{2}\sum_{\epsilon_{1},\epsilon_{2},\epsilon_{3} = \pm 1} \epsilon_{1}\epsilon_{2}\epsilon_{3}\tilde{\Psi}_{p_{1}p_{2}p_{3}}^{p_{1}p_{2}p_{3}(1+\sum_{j}\epsilon_{j}/p_{j})}(1/k), \label{eqn:linSumModular}
\end{gather}
where $q$ in Equation \ref{eqn:modular} is given by $e^{2 \pi i /k }$.

Now, let us restrict ourselves to $(p_{1},p_{2},p_{3}) = (2,5,7)$. First of all, $p_{1}=2,p_{2}=5,p_{3}=7$ are relatively prime, so $\Sigma(2,5,7)$ is a homology sphere. Next, $1/p_{1}+1/p_{2}+1/p_{3} < 1$, so we can write the WRT invariant as a linear sum of mock modular forms:

\begin{align}
e^{\frac{2 \pi i}{k}(\frac{\phi(2,5,7)}{4}-\frac{1}{2})}(e^{\frac{2 \pi i}{k}} -1) \tau_{k}(\Sigma(2,5,7)) &= \frac{1}{2}\tilde{\Psi}_{70}^{(1,1,1)}(1/k) \nonumber \\
&= \frac{1}{2}(\tilde{\Psi}_{70}^{11} - \tilde{\Psi}_{70}^{31}  - \tilde{\Psi}_{70}^{39} + \tilde{\Psi}_{70}^{59})(1/k),
\label{eqn:WRTinvariantModularDecomposition}
\end{align}
where $\phi(2,5,7) = -\frac{19}{70}$. From Equation \ref{eqn:WRTandCSpartition} and \ref{eqn:WRTinvariantModularDecomposition}, we can explicitly write the exact Chern-Simons partition function $Z_{CS}(\Sigma(2,5,7))$ as follows:

\begin{equation}
Z_{CS}(\Sigma(2,5,7)) = \frac{1}{i q^{\phi(2,5,7)/4}\sqrt{8k}}(\tilde{\Psi}_{70}^{11} - \tilde{\Psi}_{70}^{31}  - \tilde{\Psi}_{70}^{39} + \tilde{\Psi}_{70}^{59})(1/k).
\label{eqn:CSpartition}
\end{equation} 

\section{Asymptotics of $Z_{CS}(\Sigma(2,5,7))$}
\label{sec:asymptotics}
Before proceeding to the Borel transform and resummation of the exact partition function, let us briefly consider the its asymptotics in the large $k$ limit. This can be most easily done by considering the ``mock modular'' property of mock modular forms:
\begin{gather}
\tilde{\Psi}_{p}^{a}(q) = -\sqrt{\frac{k}{i}} \sum_{b=1}^{p-1}\sqrt{\frac{2}{p}}\sin{\frac{\pi a b}{p}}\tilde{\Psi}_{p}^{b}(e^{-2 \pi i k}) + \sum_{n \geq 0} \frac{L(-2n,\psi_{2p}^{a})}{n!}\bigg(\frac{\pi i}{2 p k}\bigg)^{n}, \label{eqn:mockmodular} \\
\text{where} \quad L(-n,\psi_{2p}^{a}) = -\frac{(2p)^{n}}{n+1}\sum_{m=1}^{2p}\psi_{2p}^{a}(m)B_{n+1}\bigg(\frac{m}{2p}\bigg),
\end{gather}
and $B_{n+1}$ stands for the $(n+1)$-th Bernoulli polynomial. For integer values of $k$, 
$$\tilde{\Psi}^{b}_{p}(e^{-2 \pi i k}) = (1-\tfrac{b}{p})e^{-\frac{2 \pi i k b^{2}}{2p}},$$ and in large $k$ limit, we may consider the second summation in Equation \ref{eqn:mockmodular} as ``perturbative'' contributions, while the first summation standing for ``non-perturbative'' contributions. Therefore, the asymptotics of $Z_{CS}(\Sigma(2,5,7))$ can be written as ($p=70$, below):
\begin{multline}
i q^{-19/280}\sqrt{8k}Z_{CS}(\Sigma(2,5,7)) = \\
 -\sqrt{\frac{k}{i}} \sum_{b=1}^{70-1} \sqrt{\frac{2}{70}}\bigg( \sin\frac{11b \pi}{70} - \sin\frac{31b \pi}{70} - \sin\frac{39b \pi}{70} + \sin\frac{59b \pi}{70}\bigg)(1-\tfrac{b}{p})e^{-\frac{2 \pi i k b^{2}}{4p}} \\
+ i q^{-19/280}\sqrt{8k}Z_{\text{pert}}(1/k),
\label{eqn:asymptotics}
\end{multline}
where the perturbative contributions $i q^{-19/280}\sqrt{8k}Z_{\text{pert}}(1/k)$ can be explicitly written as:
\begin{gather}
Z_{\text{pert}}(1/k) = Z^{11}_{\text{pert}}(1/k)-Z^{31}_{\text{pert}}(1/k)-Z^{39}_{\text{pert}}(1/k)+Z^{59}_{\text{pert}}(1/k), \nonumber \\
\text{where} \quad i \sqrt{8} q^{-19/280} Z^{a}_{\text{pert}}(1/k) = \sum_{n \geq 0} \frac{b_{n}^{a}}{k^{n+1/2}} \quad \text{for} \quad a = 11, 31, 39, 59 \nonumber \\
\text{and} \quad b_{n}^{a} = \frac{L(-2n,\psi_{2p}^{a})}{n!}\bigg(\frac{\pi i}{2p}\bigg)^{n}.
\end{gather}

One can easily see that the sum $\big( \sin\frac{11b \pi}{70} - \sin\frac{31b \pi}{70} - \sin\frac{39b \pi}{70} + \sin\frac{59b \pi}{70}\big)$ in Equation \ref{eqn:asymptotics} is nonzero if and only if $b$ is not divisible by $2,5$ or $7$. We will later see that these $b$'s correspond to the positions of the poles in the Borel plane.

\section{Resurgence analysis of $Z_{CS}(\Sigma(2,5,7))$}
In this section, we perform a resurgence analysis of the partition function and decompose $Z_{CS}(M(2,5,7))$ into the homological blocks:
$$Z_{CS}(\Sigma(2,5,7)) = \sum_{\alpha} n_{\alpha}e^{2 \pi i k CS(\alpha)} Z_{\alpha},$$
where $\alpha$ runs over the abelian/reducible flat connections. Since $Z_{\alpha}$ gets contributions from both the abelian/reducble flat connection $\alpha$ and the irreducible flat connections, it is necessary to study how the contributions from the irreducible flat connections regroup themselves into the homological blocks. We accomplish the goal in three steps. First, we study the Borel transform and resummation of the partition function and identify the contributions from the irreducible flat connections. Then, the contributions from the irreducible flat connections are shown to enter in the homological blocks via Stokes monodromy coefficients.

\subsection{Borel transform and resummation of $Z_{CS}(M(2,5,7))$}
Recall that the perturbative contributions $Z^{a}_{\text{pert}}(1/k)$ have the following asymptotics:
\begin{equation}
i\sqrt{8}q^{\phi(2,5,7)/4}Z^{a}_{\text{pert}}(1/k) = \sum_{n \geq 0} \frac{b^{a}_{n}}{k^{(n+1/2)}}.
\end{equation}
Now, consider its Borel transform:
\begin{align}
BZ^{a}_{\text{pert}}(\zeta) &= \sum_{n \geq 1} \frac{b^{a}_{n}}{\Gamma(n+1/2)} \zeta^{n-1/2} \\
&= \frac{1}{\sqrt{\zeta}} \sum_{n \geq 0} b^{a}_{n}\frac{4^{n}}{\sqrt{\pi}} \frac{n!}{(2n)!}\zeta^{n} \quad \bigg(\because \Gamma(n+1/2) = \frac{\sqrt{\pi}}{4^{n}}\frac{(2n)!}{n!} \bigg) \\
&= \frac{1}{\sqrt{\pi \zeta}} \sum_{n \geq 0} c^{a}_{n} \frac{n!}{(2n)!} z^{2n}, \quad \text{where} \quad z = \sqrt{\frac{2 \pi i}{p}\zeta}.
\end{align}
In the last equality, we have simply changed the variable from $\zeta$ to $z$ and absorbed all other factors into the coefficients $c_{n}^{a}$.

Although the coefficients $c^{a}_{n}$ only appear in the perturbative piece of the partition function, we can recover the exact partition function from them. Let us first consider generating functions which package the coefficients $c^{a}_{n}$:
$$\frac{\sinh((p-a)z)}{\sinh(pz)} = \sum_{n \geq 0} c^{a}_{n} \frac{n!}{(2n)!} z^{2n} = \sum_{n \geq 0} \psi_{2p}^{a} e^{-nz}.$$ 
Now we can write the mock modular forms in an integral from, using these generating functions:
\begin{gather}
\frac{\sinh(p-a)\eta}{\sinh p \eta} = \sum_{n \geq 0} \psi_{2p}^{a}(n)e^{-n \eta} \\
\Rightarrow \quad \int_{i \mathbb{R} + \epsilon} d \eta \frac{\sinh(p-a)\zeta}{\sinh p \eta} e^{-\frac{k p \eta^{2}}{2 \pi i}} = \int_{i \mathbb{R}+\epsilon} d \eta \sum_{n \geq 0} \psi_{2p}^{a}(n)e^{-n \eta} e^{-\frac{k p \eta^{2}}{2 \pi i}} \label{eqn:Borel1} \\
\Rightarrow  \quad \int_{i \mathbb{R} + \epsilon} d \eta \frac{\sinh(p-a)\eta}{\sinh p \eta} e^{-\frac{k p \eta^{2}}{2 \pi i}} = \sqrt{\frac{2 \pi^{2} i}{p}} \frac{1}{\sqrt{k}} \tilde{\Psi}_{p}^{a}(q). \label{eqn:Borel2}
\end{gather}
In the second line, the integral is taken along a line $Re[\eta] = \epsilon > 0$, where the integral converges, and the third line is simply a Gaussian integral. The change of variables
$$ \zeta = \frac{p \eta^{2}}{2 \pi i}$$
alters the integration contour from a single line to the union of two rays from the origin, $i e^{i \delta} \mathbb{R}_{+}$ and $i e^{-i \delta} \mathbb{R}_{+}$. In sum, 
\begin{equation}
\frac{1}{\sqrt{k}} \tilde{\Psi}_{p}^{a}(q) = \frac{1}{2}\bigg(\int_{i e^{i \delta} \mathbb{R}_{+}} + \int_{i e^{-i \delta} \mathbb{R}_{+}} \bigg) \frac{d \zeta}{\sqrt{\pi \zeta}} \frac{\sinh \bigg( (p-a)\sqrt{\frac{2 \pi i \zeta}{p}}\bigg)}{\sinh \bigg(p \sqrt{\frac{2 \pi i \zeta}{p}}\bigg)} e^{-k\zeta}. \label{eqn:BorelZeta}
\end{equation}

Thus we have recovered the entire mock modular form from its perturbative expansion. Since the partition function is a linear sum of mock modular forms, this implies that the Borel resummation of $BZ_{\text{pert}}$ will return the exact partition function. Furthermore, the poles of generating functions $\sinh((p-a)z)/\sinh(pz)$ encodes the information of the non-perturbative contributions, as we exhibit below.

\begin{figure} [htb]
\centering
\includegraphics{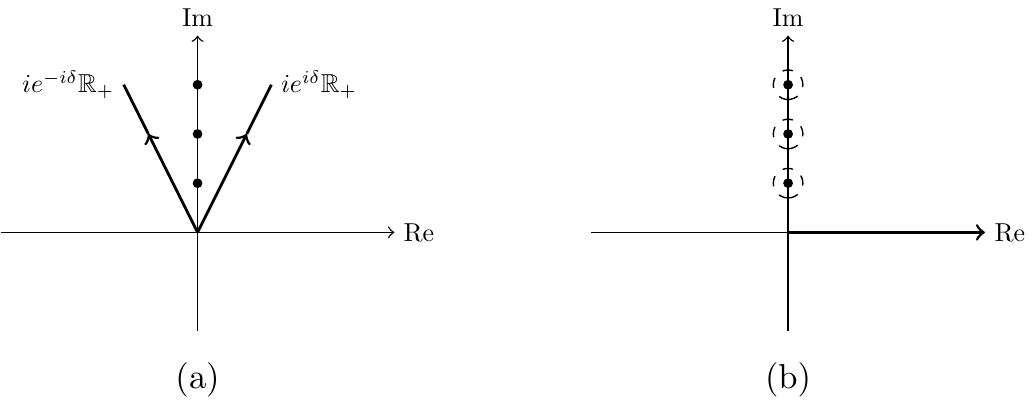}
\caption{(a) An integration contour in the $\zeta$-plane, made of two rays from the origin. Dots represent the poles. (b) An equivalent integration contour. The contribution from the integration along the real axis must be doubled.}
\label{fig:integrationContour}
\end{figure}

First of all, since $Z_{CS}(\Sigma(2,5,7)) \sim (\tilde{\Psi}_{70}^{11} - \tilde{\Psi}_{70}^{31}  - \tilde{\Psi}_{70}^{39} + \tilde{\Psi}_{70}^{59})(q)$, the Borel transform of $Z_{\text{pert}}$ is given by:
\begin{equation}
\frac{\sinh(59\eta)-\sinh(39\eta)-\sinh(31\eta)+\sinh(11\eta)}{\sinh(70\eta)} = \frac{4\sinh(35\eta)\sinh(14\eta)\sinh(10\eta)}{\sinh(70\eta)},
\label{eqn:psiBorel}
\end{equation}
Note that the RHS of Equation \ref{eqn:psiBorel} has only simple poles at $\eta = n \pi i/ 70$ for $n$ non-divisible by 2, 5, or 7. In particular, the poles are aligned on the imaginary axis, so we choose the same integration contours as in Equation \ref{eqn:Borel1} - \ref{eqn:BorelZeta}. The Borel resummation of Equation \ref{eqn:psiBorel} is then the average of Borel sums along the two rays depicted in Figure \ref{fig:integrationContour}(a):
\begin{equation}
Z_{CS}(\Sigma(2,5,7)) = \frac{1}{2} \bigg[ S_{\frac{\pi}{2} - \delta}Z_{\text{pert}}(1/k) + S_{\frac{\pi}{2} + \delta}Z_{\text{pert}}(1/k) \bigg].
\label{eqn:Borelsums}
\end{equation}

To evaluate the RHS of Equation \ref{eqn:Borelsums}, we integrate along an equivalent contour in Figure \ref{fig:integrationContour}(b). Note that as we change to the contour in Figure \ref{fig:integrationContour}(b), a Stokes ray $i e^{-i \delta}\mathbb{R}_{+}$ has crossed the poles on the imaginary axis, towards the positive real axis. As a reult, the poles contribute to the Borel sums with residues, which is precisely a Stokes phenomenon. Since each pole is located at $\eta = n \pi i / 70$, its residue includes a factor of $e^{-k \zeta} = e^{-k \frac{70 \eta^{2}}{2 \pi i}} = e^{2 \pi i k (- \frac{n^{2}}{280})}$. Shortly, we will exhibit that these factors precisely correspond to the Chern-Simons instanton actions, so let us regroup the poles ($n$ modulo 140) by their instanton actions:

\begin{itemize}
\item $n = 9, 19, 51, 61, 79, 89, 121, 131$, for which $CS = -\frac{9^{2}}{280}$ and residues $\{1,1,1,1,-1,-1,-1,-1\}$ with overall factor $\frac{i}{35}(\cos \frac{3 \pi}{35} - \sin \frac{\pi}{70})$.
\item $n = 3, 17, 53, 67, 73, 87, 123, 137$, for which $CS = -\frac{3^{2}}{280}$ and residues $\{ -1, -1, -1,-1,1,1,1,1\}$ with overall factor $\frac{i}{35}(\cos \frac{\pi}{35} + \cos \frac{6 \pi}{35})$.
\item $n = 23, 33, 37, 47, 93, 103, 107, 117$, for which $CS = -\frac{23^{2}}{280}$ and residues $\{1, 1, 1, 1,-1,-1,-1,-1\}$ with overall factor $\frac{i}{35}(\cos \frac{4 \pi}{35} + \sin \frac{13 \pi}{70})$.
\item $n = 13, 27, 43, 57, 83, 97, 113, 127$, for which $CS = -\frac{13^{2}}{280}$ and residues $\{ -1, -1, -1, -1,1,1,1,1\}$ with overall factor $\frac{i}{35}(\sin \frac{3 \pi}{70} + \sin \frac{17 \pi}{70})$.
\item $n = 11, 31, 39, 59, 81, 101, 109, 129$, for which $CS = -\frac{11^{2}}{280}$ and residues $\{1, -1, -1, 1, -1,1,1,-1\}$ with overall factor $\frac{i}{35}(\cos \frac{8 \pi}{35} + \sin \frac{9 \pi}{70})$.
\item $n = 1, 29, 41, 69, 71, 99, 111, 139$, for which $CS = -\frac{1^{2}}{280}$ and residues $\{1,-1,-1,1,-1,1,1,-1\}$ with overall factor $\frac{i}{35}(\cos \frac{2 \pi}{35} - \sin \frac{11 \pi}{70})$.
\end{itemize}

The top four groups of poles correspond to the four irreducible $SU(2)$ flat connections, while the remaining two correspond to the complex flat connections. To see this, first consider the moduli space of flat connections $\mathcal{M}_{\text{flat}}(\Sigma(2,5,7), SL(2,\mathbb{C}))$. Since $\Sigma(2,5,7)$ is a homology 3-sphere, it has only one abelian flat connection $\alpha_{0}$, which is trivial. Next, there are total $\frac{(2-1)(5-1)(7-1)}{4} = 6$ irreducible $SL(2,\mathbb{C})$ flat connections, four of which are conjugate to $SU(2)$ and the remaining two are ``complex'' (conjugate to $SL(2,\mathbb{R})$) \cite{KitanoYamaguchi,BodenCurtis,FintushelStern}. To compute their Chern-Simons instanton actions, we characterize all six flat connections by their ``rotation angles,'' which we will briefly explain here. Consider the following presentation of the fundamental group of $\Sigma(2,5,7)$.
\begin{equation}
\pi_{1}(\Sigma(2,5,7)) = \langle x_{1},x_{2},x_{3},h \, | \, h \, \text{central}, x_{1}^{2} = h^{-1}, x_{2}^{5} = h^{-9}, x_{3}^{7} = h^{-5}, x_{1}x_{2}x_{3} = h^{-3} \rangle.
\label{eqn:fundPresentation}
\end{equation}
When a representation $\alpha: \pi_{1}(\Sigma(2,5,7)) \rightarrow SL(2,\mathbb{C})$ is conjugate in $SU(2)$, $\alpha(h)$ is equal to $\pm 1$, and the conjugacy classes of $\alpha(x_{j})$ can be represented in the form $\bigl(\begin{smallmatrix}
\lambda_{j} & 0 \\ 0 & \lambda_{j}^{-1}
\end{smallmatrix} \bigr)$ for some $| \lambda_{j} | = 1$. There are four triples $(\lambda_{1}, \lambda_{2}, \lambda_{3})$ satisfying the relations in Equation \ref{eqn:fundPresentation}:
\begin{equation}
(l_{1},l_{2},l_{3}) = (1,1,3), \, (1,3,1), \, (1,3,3), \, (1,3,5) \quad \text{where} \quad \lambda_{j} = e^{\pi i l_{j} / p_{j}}. \label{eqn:flatCon}
\end{equation} 
Each triple corresponds to one of the four irreducible $SU(2)$ flat connections, which we will call $\alpha_{1}, \alpha_{2}, \alpha_{3}$ and $\alpha_{4}$. From the rotation angles of an irreducible flat connection $A$, we can read off its Chern-Simons instanton action:
\begin{gather}
CS(A) = -\frac{p_{1}p_{2}p_{3}}{4}(1+\sum_{i}l_{j}/p_{j})^{2} \nonumber \\
\Rightarrow \quad CS(\alpha_{1}) = -\frac{9^{2}}{280}, \quad CS(\alpha_{2}) = -\frac{3^{2}}{280}, \quad CS(\alpha_{3}) = -\frac{23^{2}}{280}, \quad CS(\alpha_{4}) = -\frac{13^{2}}{280}, \label{eqn:CSinstanton}
\end{gather}
which is in agreement with the instanton actions of the poles in the Borel plane. Likewise, one can compute the Chern-Simons instanton actions of the two complex flat connections $\alpha_{5}$ and $\alpha_{6}$, 
$$CS(\alpha_{5}) = -\frac{11^{2}}{280}, \quad CS(\alpha_{6}) = -\frac{1^{2}}{280}.$$

Now, let us sum the residues to reproduce the non-perturbative contributions in Equation \ref{eqn:asymptotics}. When $k$ is an integer, the residues from the poles $SU(2)$ connection $\alpha_{1}$ are summed into:
\begin{multline}
\frac{i}{35}(\cos \frac{3 \pi}{35} - \sin \frac{\pi}{70}) e^{-2 \pi i k \frac{9^{2}}{280}} \bigg[ \sum_{n \equiv \pm 9 \, (mod \, 140)} \pm 1 \, + \sum_{n \equiv \pm 19 \, (mod \, 140)} \pm 1 \\
+ \sum_{n \equiv \pm 51\, (mod \, 140)} \pm 1 \, + \sum_{n \equiv \pm 61\, (mod \, 140)} \pm 1 \bigg].
\label{eqn:alpha1Contribution}
\end{multline}
Via zeta-function regularization $\sum_{n \equiv \pm a \, (mod \, 2p)} \pm 1 = 1 - \frac{a}{p}$, we can rewrite Equation \ref{eqn:alpha1Contribution} as follows:
\begin{gather*}
\frac{i}{35}(\cos \frac{3 \pi}{35} - \sin \frac{\pi}{70}) e^{-2 \pi i k \frac{9^{2}}{280}}\bigg( (1-\frac{9}{70}) + (1-\frac{19}{70}) + (1-\frac{51}{70}) + (1-\frac{61}{70}) \bigg) \\
= \frac{2i}{35}(\cos \frac{3 \pi}{35} - \sin \frac{\pi}{70}) e^{-2 \pi i k \frac{9^{2}}{240}} = n_{\alpha_{1}}Z_{\text{pert}}^{\alpha_{1}}e^{2 \pi i k CS(\alpha_{1})},
\end{gather*}
where $n_{\alpha_{1}}$ is the trans-series parameter. Similarly for connections $\alpha_{2}, \alpha_{3}$ and $\alpha_{4}$, 
\begin{itemize}
\item $n_{\alpha_{2}}Z_{\text{pert}}^{\alpha_{2}}e^{2 \pi i k CS(\alpha_{2})} = -\frac{2i}{35}(\cos \frac{\pi}{35} + \cos \frac{6 \pi}{35})e^{-2 \pi i k \frac{3^{2}}{280}}$.
\item $n_{\alpha_{3}}Z_{\text{pert}}^{\alpha_{3}}e^{2 \pi i k CS(\alpha_{3})} = \frac{2i}{35}(\cos \frac{4 \pi}{35} + \sin \frac{13 \pi}{70})e^{-2 \pi i k \frac{23^{2}}{280}}$.
\item$n_{\alpha_{4}}Z_{\text{pert}}^{\alpha_{4}}e^{2 \pi i k CS(\alpha_{4})} = -\frac{2i}{35}(\sin \frac{3 \pi}{70} + \sin \frac{17 \pi}{70})e^{-2 \pi i k \frac{13^{2}}{280}}$.
\end{itemize}
And the contributions from the two complex connections vanish. Notice that the poles grouped by their instanton actions correspond to the $b$'s with non-vanishing contributions in Equation \ref{eqn:asymptotics}. Furthermore, the sum of residues is proportional to the sum $\big( \sin\frac{11b \pi}{70} - \sin\frac{31b \pi}{70} - \sin\frac{39b \pi}{70} + \sin\frac{59b \pi}{70}\big)$ at each $b$, so the Borel sum correctly captures the non-perturbative contributions to the exact partition function. 

\section{Homological block decomposition of $Z_{CS}(\Sigma(2,5,7))$ and the modular transform}
We conclude this paper by writing the partition function in a categorification-friendly form, as was advertised in Equation \ref{eqn:abelianDecomposition}. To summarize, we started with the exact partition function $Z_{CS}(\Sigma(2,5,7))$, considered its perturbative expansion and performed a Borel resummation. Although our example is a homology 3-sphere and has only one abelian flat connection, more generally the Borel sum results in a decomposition into homological blocks \cite{GMP}:
$$Z_{CS}(\Sigma(2,5,7)) = \sum_{a} e^{2 \pi i CS_{a}}Z_{a},$$
where the summation runs over abelian flat connections. Each ``homological block'' $Z_{a}$ gets contributions from both the abelian flat connection $a$ and the irreducible $SU(2)$ flat connections. How the irreducible flat connections regroup themselves into each homological block is encoded in the Stokes monodromy coefficients as follows:
\begin{gather}
Z_{CS}(\Sigma(2,5,7)) = \frac{1}{2}\bigg[S_{\frac{\pi}{2}-\epsilon}Z_{\text{pert}}(k) + S_{\frac{\pi}{2}+\epsilon}Z_{\text{pert}}(k) \bigg] = Z^{\alpha_{0}}_{\text{pert}} + \frac{1}{2} \sum_{\tilde{\beta}}m_{\tilde{\beta}}^{(\alpha_{0},0)}e^{2 \pi i k S_{\tilde{\beta}}}Z^{\beta}_{\text{pert}} \nonumber \\
=\sum_{\tilde{\beta}} n_{\tilde{\beta},0}e^{2 \pi i k S_{\tilde{\beta}}}Z^{\beta}_{\text{pert}},  \quad \text{where} \quad n_{\tilde{\beta}} = \begin{cases} 1 & \tilde{\beta}=(\alpha_{0},0) \\ \frac{1}{2}m_{\tilde{\beta}}^{(\alpha_{0},0)} & \text{otherwise.} \end{cases} \label{eqn:trans-seriesCoeff}
\end{gather}
\begin{equation}
m_{\tilde{\beta}}^{(\alpha_{0},0)} = \begin{cases}
1 & \tilde{\beta} = (\alpha_{1}, -n^{2}/280), \quad \text{for} \quad n = 9, 19, 51, 61 \quad (\text{mod} \, 140) \\
-1 & \tilde{\beta} = (\alpha_{1}, -n^{2}/280), \quad \text{for} \quad n = 79, 89, 121, 131 \quad (\text{mod} \, 140) \\
1 & \tilde{\beta} = (\alpha_{2}, -n^{2}/280), \quad \text{for} \quad n = 73, 87, 123, 137 \quad (\text{mod} \, 140) \\
-1 & \tilde{\beta} = (\alpha_{2}, -n^{2}/280), \quad \text{for} \quad n = 3, 17, 53, 67 \quad (\text{mod} \, 140) \\
1 & \tilde{\beta} = (\alpha_{3}, -n^{2}/280), \quad \text{for} \quad n = 23, 33, 37, 47 \quad (\text{mod} \, 140) \\
-1 & \tilde{\beta} = (\alpha_{3}, -n^{2}/280), \quad \text{for} \quad n = 93, 103, 107, 117 \quad (\text{mod} \, 140) \\
1 & \tilde{\beta} = (\alpha_{4}, -n^{2}/280), \quad \text{for} \quad n = 83, 97, 113, 127 \quad (\text{mod} \, 140) \\
-1 & \tilde{\beta} = (\alpha_{4}, -n^{2}/280), \quad \text{for} \quad n = 13, 27, 43, 57 \quad (\text{mod} \, 140) \\
1 & \tilde{\beta} = (\alpha_{5}, -n^{2}/280), \quad \text{for} \quad n = 11, 59, 101, 109 \quad (\text{mod} \, 140) \\
-1 & \tilde{\beta} = (\alpha_{5}, -n^{2}/280), \quad \text{for} \quad n = 31, 39, 81, 129 \quad (\text{mod} \, 140) \\
1 & \tilde{\beta} = (\alpha_{6}, -n^{2}/280), \quad \text{for} \quad n = 1, 69, 99, 111 \quad (\text{mod} \, 140) \\
-1 & \tilde{\beta} = (\alpha_{6}, -n^{2}/280), \quad \text{for} \quad n = 29, 41, 71, 139 \quad (\text{mod} \, 140) \end{cases}
\end{equation}
The formula \ref{eqn:trans-seriesCoeff} holds for any Seifert manifold with three singular fibers (which includes our example $\Sigma(2,5,7)$) \cite{CostinGaroufalidis}. In \cite{GPV}, it was conjectured that there is a ``modular transform'' of the homological blocks $Z_{a}$, which turns it into a ``categorification-friendly'' form. Namely,
$$Z_{a} = \frac{1}{i \sqrt{2k}} \sum_{b} S_{ab} \hat{Z}_{b},$$
for some $k$-independent $S_{ab}$. Above, $b$ runs over the abelian flat connections, and each $\hat{Z}_{b}$ is an element of $q^{\Delta_{b}}\mathbb{Z}[[q]]$ for some $\Delta_{b} \in \mathbb{Q}$. Suppose the exact partition function is a linear sum of mock modular forms, and there are multiple abelian flat connections. Then, a homological block decomposition regroups the mock modular forms (see \cite{GPV} for examples.) In our example, however, there is only one abelian flat connection $\alpha_{0}$, because $\Sigma(2,5,7)$ is a homology sphere. Therefore, it suffices to find $\hat{Z}_{\alpha_{0}}$ which is an element of $q^{\Delta_{\alpha_{0}}} \mathbb{Z}[[q]]$. From the exact partition function
$$i q^{\phi(2,5,7)/4}\sqrt{2k}Z_{CS}(\Sigma(2,5,7)) = \frac{1}{2}(\tilde{\Psi}_{70}^{11} - \tilde{\Psi}_{70}^{31}  - \tilde{\Psi}_{70}^{39} + \tilde{\Psi}_{70}^{59})(q),$$
and the definition of the mock modular forms
$$\tilde{\Psi}_{p}^{a}(q) = \sum_{n \geq 0} \psi_{2p}^{a}q^{n^{2}/4p},$$
we can easily see that the partition function is an element of $q^{121/280}\mathbb{Z}[[q]]$. Thus,
$$\hat{Z}_{\alpha_{0}} = q^{1/2}(1 - q^{3} - q^{5} + q^{12} + \cdots) \quad \text{and} \quad S_{\alpha_{0}\alpha_{0}} = \frac{1}{2}.$$

\acknowledgments{The author is deeply indebted to Sergei Gukov for his suggestions and invaluable discussions.

The work is funded in part by the DOE Grant DE-SC0011632 and the Walter Burke Institute for Theoretical Physics, and also by the Samsung Scholarship.}

\newpage

\bibliographystyle{JHEP_TD}
\bibliography{Example2}

\end{document}